\documentclass[conference]{IEEEtran}

\IEEEoverridecommandlockouts
\usepackage{cite}
\usepackage{amsmath,amssymb,amsfonts}
\usepackage{graphicx}
\usepackage{flushend}  

\usepackage{xcolor}
\usepackage{booktabs}
\usepackage{textcomp}
\usepackage[colorlinks=true, linkcolor=blue, urlcolor=blue, citecolor=blue]{hyperref}
\usepackage{url}
\graphicspath{{./Figures/}}

\def\BibTeX{{\rm B\kern-.05em{\sc i\kern-.025em b}\kern-.08em
    T\kern-.1667em\lower.7ex\hbox{E}\kern-.125emX}}
    
\RequirePackage{silence}
\definecolor{acceptedbg}{RGB}{232,245,233}
\definecolor{acceptedborder}{RGB}{46,125,50}

\newcommand{\AcceptedPaperNote}{%
\begingroup
\setlength{\fboxsep}{6pt}%
\setlength{\fboxrule}{0.5pt}%
\noindent\fcolorbox{acceptedborder}{acceptedbg}{%
\parbox{\dimexpr\textwidth-2\fboxsep-2\fboxrule\relax}{%
\normalfont\footnotesize
\textbf{Acceptance note:}
This work-in-progress paper has been accepted for presentation at the
\textit{IEEE WoWMoM 2026 Work-in-Progress (WiP) track}.
}%
}%
\par
\endgroup
}

\begin{document}

\title{%
\AcceptedPaperNote

WIP: Environment-Aware Indoor LoRaWAN Ranging Using Path Loss Model Inversion and Adaptive RSSI Filtering \\

\thanks{This work was partly supported by the German Academic Exchange Service (DAAD) under Grant 57652455. Additional funding was provided by the German Research Foundation (DFG) under Grant 425868829 as part of Priority Program SPP2199: Scalable Interaction Paradigms for Pervasive Computing Environments.}
}

\author{
\IEEEauthorblockN{Nahshon Mokua Obiri and Kristof Van Laerhoven}
\IEEEauthorblockA{\textit{Department of Electrical Engineering and Computer Science, University of Siegen} \\
E-mails: \{nahshon.obiri@student.uni-siegen.de, kvl@eti.uni-siegen.de\}}
}

\maketitle

\begin{abstract}
Achieving sub-10\,m indoor ranging with LoRaWAN is challenging because multipath, human blockage, and micro-climate dynamics induce non-stationary attenuation in received signal strength indicator (RSSI) measurements. We present a lightweight, interpretable, site-calibrated pipeline that couples an environment-aware multi-wall path loss model with a forward-only, innovation-driven Kalman prefilter for RSSI. The model augments distance and wall terms with frequency, signal-to-noise ratio (SNR), and co-located environmental covariates (temperature, relative humidity, carbon dioxide, particulate matter, and barometric pressure), and is inverted deterministically for distance estimation. On a one-year single-gateway office dataset comprising over $2$ million uplinks, the approach attains a mean absolute error (MAE) of $4.74\,\mathrm{m}$ and a root mean square error (RMSE) of $6.76\,\mathrm{m}$ in distance estimation, improving over a structure-only COST-231 multi-wall baseline ($12.07\,\mathrm{m}$ MAE) and its environment-augmented variant without filtering ($7.76\,\mathrm{m}$ MAE). Filtering reduces RSSI volatility from $10.33$ to $5.43\,\mathrm{dB}$ and lowers the path loss RMSE from $8.09$ to $5.35\,\mathrm{dB}$, while increasing $R^2$ from $0.82$ to $0.89$. The result is a single-anchor LoRaWAN ranging method with an $O(1)$ per-packet cost that is stable, interpretable, and practical within a calibrated indoor deployment, providing a useful building block for future multi-gateway localization and a benchmark for indoor LoRaWAN ranging.
\end{abstract}



\section{Introduction}

LoRaWAN is not a substitute for high-precision indoor positioning technologies, but it remains attractive in indoor Internet of Things (IoT) deployments where long communication range, sparse infrastructure, low-power operation, deep penetration, and reuse of existing uplink traffic matter \cite{fahmidaRTPLRealTimeCommunication2024, obiriSurveyLoRaWANIntegratedWearable2024}. This is especially relevant in large indoor spaces and building-scale deployments, where recent LoRa localization work has considered settings such as warehouses, airport terminals, sports centers, and museum halls \cite{guoILLOCInHallLocalization2022}. Within this operating regime, achieving sub-10\,m ranging from a single gateway remains challenging. The received signal strength indicator (RSSI) is shaped by multipath, human blockage, and micro-climate dynamics such as temperature or humidity cycles, which induce non-stationary attenuation and shadowing \cite{syazreenahmadRecentAdvancesWSNBased2024, weiRSSIbasedLocationFingerprint2024}. As a result, single-gateway RSSI ranging methods typically report $8\!-\!20\,\mathrm{m}$ error \cite{rathnayakeRSSIMachineLearningBased2023}, which limits their utility as a standalone localization mechanism but still leaves room for a robust ranging primitive that can support multi-gateway localization.

Classical propagation models such as the COST‑231 multi‑wall model (MWM) \cite{europeancommissionCOSTAction2311999}  provide a physically grounded starting point but do not explain time‑varying environmental influences \cite{azevedoCriticalReviewPropagation2024}. Empirical work has shown that temperature and humidity affect attenuation at $868\,\mathrm{MHz}$ via water‑vapor absorption \cite{cattaniExperimentalEvaluationReliability2017, lavdasEffectTemperatureHumidity2021}, while occupancy alters diffraction and scattering \cite{grubelDenseIndoorSensor2022b}. Treating these effects as stationary log‑normal shadowing results in a systematic distance bias.

Data-driven approaches, such as fingerprinting, random forests, or deep neural networks, can achieve strong indoor accuracy by learning complex interactions among RSSI, signal-to-noise ratio (SNR), and the environmental or spatial context \cite{rathnayakeRSSIMachineLearningBased2023, voAdvancePathLoss2024}. Their deployment, however, typically depends on site-specific retraining, labeled calibration data, or radiomap maintenance, and their internal decision structure is often less transparent during diagnosis. This is particularly relevant indoors, where recent reviews show that empirical LoRa propagation models can lose accuracy outside the environments in which they were fitted. At the same time, richer environmental characterization usually improves performance at the cost of additional calibration effort. Our objective is therefore not building-agnostic zero-calibration transfer, but a lightweight and interpretable ranging model that keeps this calibration burden explicit and bounded.

We target physics-guided, interpretable single-gateway distance estimation under indoor variability, with $O(1)$ per-packet cost and a calibration burden that remains operationally manageable, by combining an environment-aware multi-wall path loss model with adaptive Kalman filtering of RSSI. We first extend the baseline \textbf{MWM} by adding frequency, SNR, and co-located environmental covariates (temperature, relative humidity, carbon dioxide (CO$_2$), particulate matter (PM$_{2.5}$), and barometric pressure), yielding the environment-aware model \textbf{MWM-EP}, which we invert deterministically for range. We then stabilize the RSSI with a per-device, forward-only, innovation-driven Kalman filter that self-tunes the measurement covariance, reducing volatility before inversion \cite{kalmanNewApproachLinear1960, liaoDynamicSelfTuningMaximum2022, bultenKalmanFiltersExplained2015}. In this paper, the term \textit{adaptive} refers specifically to the online covariance update, not to the online refitting of the path loss model coefficients. The resulting inference pipeline is denoted \textbf{MWM-EP-KF}. On a one-year single-gateway office dataset with over $2$ million uplinks, this pipeline achieves sub-10\,m distance estimation accuracy. Relative to recent indoor LoRa localization approaches that increasingly rely on fingerprinting, transfer learning, or heavier radiomap construction, our emphasis is different: we aim for a lightweight, interpretable, site-calibrated ranging benchmark with bounded adaptation cost and constant per-packet inference. The contributions of this work are twofold:

\begin{itemize}
    \item We develop an environment-aware multi-wall ranging model that extends structural path loss inversion with packet-level SNR and co-located environmental covariates, achieving a mean absolute distance error of $4.74\,\mathrm{m}$ and providing a calibrated, interpretable benchmark for indoor LoRaWAN ranging.
    \item We introduce an innovation-driven Kalman prefilter for RSSI that updates the measurement-noise covariance online, reducing RSSI volatility by $43.29\%$ and improving the stability of constant-time distance inversion without requiring a fingerprint map or model retraining at inference time.
\end{itemize}

\begin{figure}[hbt!]
    \centering
    \centerline{\includegraphics[width=\linewidth]{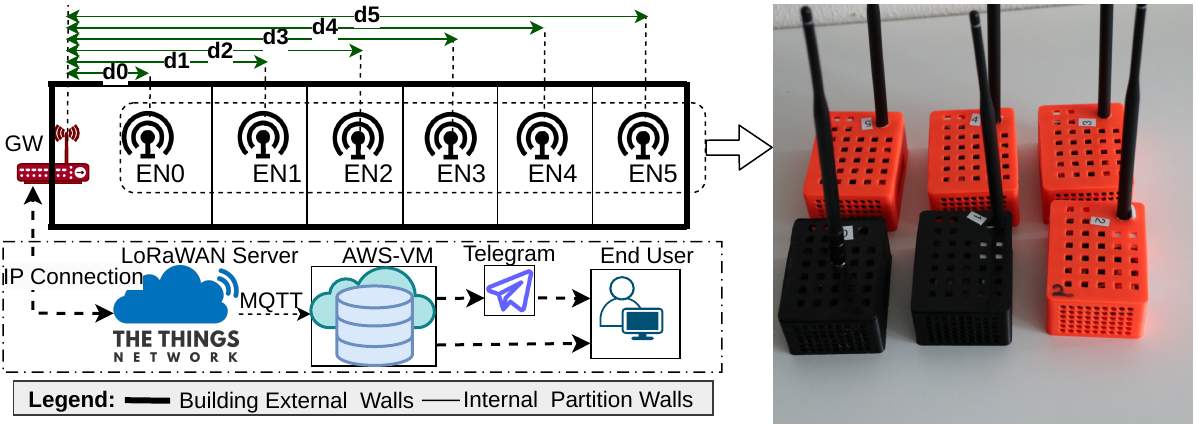}}
    \caption{\label{fig:design} 
    Experimental layout (left) and end-node prototype (right).}
\end{figure}

\section{Design and Methodology}

\subsection{Data Collection Campaign}

We collected indoor measurements spanning four typical European seasons (one year) using a single LoRaWAN gateway and six static end nodes (ENs). These end nodes were deployed on the eighth floor of an academic office building with a mix of line-of-sight (LoS) and non-line-of-sight (NLoS) links, with the network design illustrated in Fig.~\ref{fig:design}. They were positioned within a $40\,\mathrm{m}$ radius of the gateway at a common height of $0.8\,\mathrm{m}$ to induce diverse propagation conditions. Only one end node-to-gateway link was LoS, while the others were NLoS, with per-link brick and wood wall counts summarized in Table~\ref{tab:ed_wall_counts}. Ground‑truth end node-to-gateway distances and wall counts were predetermined during deployment.

\begin{table}[hbt!]
\centering
\footnotesize
\caption{Per-link wall counts (end node--to--gateway). End node labels match the network design in Fig.~\ref{fig:design}.}
\label{tab:ed_wall_counts}

{\setlength{\tabcolsep}{3pt}\footnotesize
\begin{tabular}{lcccccc}
\toprule
\textbf{End Node} & \textbf{EN0} & \textbf{EN1} & \textbf{EN2} & \textbf{EN3} & \textbf{EN4} & \textbf{EN5} \\
\midrule
Brick/Concrete walls & 0 & 1 & 0 & 1 & 0 & 2 \\
Wooden walls         & 0 & 0 & 2 & 2 & 5 & 2 \\
\bottomrule
\end{tabular}
}
\end{table}

Each end node used a vertically oriented omnidirectional antenna and a suite of co-located environmental sensors, including the Sensirion SCD41 for CO$_2$ concentration measurement, the Bosch BME280 for temperature, relative humidity, and barometric pressure, and the Sensirion SPS30 for PM$_{2.5}$ concentration. The end nodes were built on Arduino MKR WAN 1310 boards, transmitting at $868 \,\mathrm{MHz}$ with $14 \,\mathrm{dBm}$ output power within the $1\%$ duty cycle limit. Each end node sent an 18-byte uplink packet every 60\,s, embedding sensor readings and a sequence counter in a compact binary format. A Kerlink Wirnet iFemtoCell indoor gateway was configured for the EU868 band with a high receiver sensitivity of $-141 \,\mathrm{dBm}$. It forwarded uplinks to The Things Network, which relayed the data via Message Queuing Telemetry Transport (MQTT) to an Amazon Web Services (AWS)-hosted InfluxDB time-series database. Also, a continuously running Python script was deployed on an AWS Elastic Compute Cloud (EC2) instance to enable real-time, time-stamped logging of LoRaWAN link metrics (RSSI and SNR) and environmental measurements, and to raise Telegram alerts for stalled streams.

\subsection{Kalman Filtering for RSSI Refinement}
\label{subsec:kalman}

Indoor LoRaWAN RSSI exhibits short-term volatility due to transient obstructions, occupant activity, and environmental fluctuations such as heating, ventilation, and air-conditioning (HVAC)-driven humidity changes. These effects introduce irregular perturbations and occasional outliers that can destabilize direct path loss inversion. Our objective is therefore to recover, for each device, a smoothed RSSI trajectory that preserves the slowly varying attenuation trend while suppressing short-lived fluctuations. To this end, we employ a lightweight forward-only one-dimensional Kalman filter, following the general principle of innovation-driven self-tuning tracking filters \cite{liaoDynamicSelfTuningMaximum2022, bultenKalmanFiltersExplained2015}. The latent state is modeled as a random walk, and the observation model is linear, enabling recursive RSSI refinement with constant per-packet cost.

We set the process-noise covariance to $Q=0.003\,\mathrm{dB}^2$, so that the filter remains responsive to slow drift while rejecting high-frequency jitter. The initial measurement-noise covariance is set to $R_0=0.22\,\mathrm{dB}^2$, which is substantially below the raw RSSI variance $(\sigma_z^2 \approx 10.33^2\,\mathrm{dB}^2)$ and thus enforces early smoothing. To accommodate time-varying interference, the scalar measurement covariance is updated online from the normalized innovation $\nu_k=z_k-\hat{x}_{k|k-1}$ through $\alpha_k=\nu_k^2/(P_{k|k-1}+R_{k-1})$ and the exponentially smoothed recursion $R_k \leftarrow \gamma R_{k-1} + (1-\gamma)\alpha_k R_{k-1}$, with $\gamma=0.99$. To prevent instability under short-lived excursions, we clip $\alpha_k\in[0.95,1.05]$ and clamp $R_k\in[0.12,0.38]\,\mathrm{dB}^2$. This innovation-driven update modulates the effective Kalman gain, allowing the filter to remain stable under bursty conditions while retaining sensitivity to gradual channel evolution. The filter is applied independently to each device stream and does not require online estimation of full noise statistics. Importantly, this adaptation is confined to the measurement-noise covariance, and the path loss model itself remains fixed after site calibration.

\subsection{Environment-Aware Path Loss Modeling}

Natural variations in the office environment introduce slow attenuation drift that is not captured by a purely structural multi-wall model \cite{mokuaobiriComprehensiveDataDescription2025}. To keep both the regression stage and the later inversion stage notationally consistent, we express the environment-aware extension in compact packet-level form. For packet $i$ on link $\ell$, let $L_{\ell,i}$ denote the measured path loss, let $\mathbf{w}_i\in\mathbb{R}^{M}$ collect the wall descriptors with coefficient vector $\boldsymbol{\omega}\in\mathbb{R}^{M}$, and let $\mathbf{e}_i\in\mathbb{R}^{P}$ collect the environmental covariates with associated coefficient vector $\boldsymbol{\varepsilon}\in\mathbb{R}^{P}$. The resulting MWM-EP model is given by
{\small
\begin{equation}
\label{eq:pl-mw-ep}
\begin{aligned}
L_{\ell,i} =\;& \beta_0 + 10n\log_{10}\!\bigl(d_i/d_0\bigr) + 20\log_{10}(f_i)
+ \boldsymbol{\omega}^{\top}\mathbf{w}_i  \\
&+  \boldsymbol{\varepsilon}^{\top}\mathbf{e}_i + k_{\gamma}\gamma_i + \psi_i ,
\end{aligned}
\end{equation}
}
where $\beta_0$ is the site-calibrated intercept, $n$ is the path loss exponent, $d_i$ is the transmitter-receiver separation for packet $i$, and $d_0$ is the reference distance, which is set to $1\,\mathrm{m}$ in our study. Furthermore, $f_i$ is the carrier frequency, $\gamma_i$ is the gateway-reported packet-level SNR in $\mathrm{dB}$, $k_{\gamma}$ is the SNR coefficient, and $\psi_i\sim\mathcal{N}(0,\sigma_{\psi}^{2})$ captures shadowing and residual unmodeled variability. In our implementation, the wall-feature vector is $\mathbf{w}_i = [N_{i,\mathrm{brick}},\,N_{i,\mathrm{wood}}]^{\top},$ where $N_{i,\mathrm{brick}}$ and $N_{i,\mathrm{wood}}$ denote the numbers of intervening brick/concrete and wooden walls, respectively. Likewise, the environmental vector is $ \mathbf{e}_i = [T_i,\,RH_i,\,CO_{2,i},\,PM_{2.5,i},\,BP_i]^{\top},$ where $T_i$ is temperature, $RH_i$ is relative humidity, $CO_{2,i}$ is carbon-dioxide concentration, $PM_{2.5,i}$ is particulate-matter concentration, and $BP_i$ is barometric pressure.

The preprocessing and fitting pipeline, fully detailed in \cite{mokuaobiriComprehensiveDataDescription2025}, is summarized as follows: \textbf{(i)} de-duplicate and clean the uplinks; \textbf{(ii)} retain spreading factor (SF) values SF7--SF10 to avoid high-SF non-stationarities; \textbf{(iii)} apply device-wise Isolation Forest anomaly filtering ($1\%$ contamination); \textbf{(iv)} split the data into $80/20$ train/test partitions; and \textbf{(v)} estimate the parameter set $(\beta_0,\, n,\, \boldsymbol{\omega},\, \boldsymbol{\varepsilon},\, k_{\gamma})$ by ordinary least squares, with performance reported using RMSE and $R^2$ on held-out data and 5-fold cross-validation. The intercept $\beta_0$ is learned from measurements rather than being fixed to a theoretical value, allowing deployment-specific constants to be absorbed directly into the fitted model. On the same indoor dataset, as established in \cite{mokuaobiriComprehensiveDataDescription2025}, augmenting a multi-wall baseline with environmental covariates reduced the RMSE from $10.58\,\mathrm{dB}$ to $8.04\,\mathrm{dB}$ and raised $R^2$ from $0.69$ to $0.82$. This trend is consistent with the findings in \cite{gonzalez-palacioMachineLearningBasedCombinedPath2023}. Analysis of variance (ANOVA) and residual analysis in \cite{obiriStatisticalEvaluationIndoor2025a} further indicate a $42\%$ reduction in unexplained variance, motivating \eqref{eq:pl-mw-ep} and its deterministic inversion. The resulting model is therefore site-calibrated rather than building-agnostic: in the present deployment, the additional site-specific inputs are limited to per-link wall counts and the environmental measurements already produced by the sensing nodes, which keeps the calibration burden explicit and operationally modest.

\subsection{Deterministic Distance Estimation}

Given a calibrated MWM-EP model, distance estimation is obtained by algebraic inversion of \eqref{eq:pl-mw-ep}. At inference time, we use the packet-level observed path loss $L_{\ell,i}^{\mathrm{obs}}$, derived from the measured RSSI or its Kalman-filtered counterpart, and set the stochastic term $\psi_i$ to zero to obtain a deterministic estimator. The resulting deterministic distance estimate $\hat d_i$ is
{\small
\begin{equation}
\label{eq:distance-inversion}
\begin{aligned}
\hat d_i &= d_0\,10^{\eta_i/(10\hat n)},\\
\eta_i &= L_{\ell,i}^{\mathrm{obs}} -\hat\beta_0 -20\log_{10}(f_i) 
-\hat{\boldsymbol{\omega}}^{\top}\mathbf{w}_i -\hat{\boldsymbol{\varepsilon}}^{\top}\mathbf{e}_i -\hat k_{\gamma}\gamma_i ,
\end{aligned}
\end{equation}
}
where $\hat\beta_0$, $\hat n$, $\hat{\boldsymbol{\omega}}$, $\hat{\boldsymbol{\varepsilon}}$, and $\hat k_{\gamma}$ denote the calibrated parameter estimates. This closed-form inversion preserves $O(1)$ per-packet complexity while explicitly compensating for carrier frequency, wall structure, environmental state, and packet-level SNR \cite{wangWeightedHybridIndoor2025}. Although neglecting $\psi_i$ can increase error under harsh NLoS conditions \cite{voAdvancePathLoss2024}, two design choices mitigate this: \textbf{(i)} environmental and SNR terms reduce reliance on unmodeled shadowing, and \textbf{(ii)} we feed Kalman-filtered RSSI into the inversion, which lowers the effective shadowing variance and stabilizes the resulting range estimates. In a Python/NumPy/Pandas implementation on an AMD Ryzen 9 7950X 16-Core Processor, the held-out forward-only Kalman filtering stage together with path loss formation required $3.89\,\mu\mathrm{s}$ per packet on average over $415{,}907$ test packets (standard deviation: $0.06\,\mu\mathrm{s}$; $10$ runs), confirming low practical inference overhead.

\section{Evaluation Results and Discussion}

All results in this section are obtained under a held-out evaluation protocol. After de-duplication and device-wise anomaly filtering, the dataset is partitioned into disjoint training and test subsets using an $80/20$ chronological split, preserving the temporal structure of the one-year campaign. This choice follows the same rationale established in our companion study \cite{obiriEnvironmentAwareIndoorLoRaWAN2025} and is consistent with recent LoRaWAN literature \cite{gonzalez-palacioMachineLearningBasedCombinedPath2023}: environmental non-stationarity, including occupancy, HVAC cycles, and microclimate variation, materially affects propagation behavior in fixed indoor deployments, so random splitting can mask these dynamics and yield overly optimistic estimates. The path loss coefficients $(\hat\beta_0,\hat n,\hat{\boldsymbol{\omega}},\hat{\boldsymbol{\varepsilon}},\hat k_{\gamma})$ are calibrated offline on the training subset only, while the Kalman-filter hyperparameters are held fixed across all experiments. At test time, each device stream is processed independently: the raw RSSI sequence is optionally filtered sequentially by the forward-only Kalman filter, the packet-level observed path loss $L_{\ell,i}^{\mathrm{obs}}$ is formed from the raw or filtered RSSI, and \eqref{eq:distance-inversion} is evaluated once per packet using the fixed calibrated coefficients together with the packet-level wall, environmental, and SNR inputs. We report path loss accuracy using root mean square error (RMSE) and $R^2$, and ranging accuracy using RMSE, mean absolute error (MAE), median absolute error, empirical cumulative error distributions, and per-device relative error.

\subsection{Kalman Filtering Performance}

The Kalman filter reduces RSSI temporal volatility ($\sigma$) by $43.29\%$ (from $10.33\,\mathrm{dB}$ to $5.43\,\mathrm{dB}$) and mitigates error skewness from $3.68$ to $0.73$. For all six device streams, the Kalman gain $K_k$ stabilizes within approximately $20$ packets, marking the end of the filter's initial transient; the adaptive measurement covariance $R_k$ continues to exhibit only a slow, bounded evolution thereafter. This indicates a short initialization phase for the filtering stage. The improved temporal regularity translates directly to more stable ranging, with the distance-estimate RMSE dropping by $83.93\%$ from $42.04\,\mathrm{m}$ to $6.76\,\mathrm{m}$. In practice, the filtered RSSI trajectories suppress short-lived outliers, such as transient human obstructions during peak occupancy, while preserving the underlying attenuation trend.

\subsection{Path Loss Modeling Evaluation} 

Coefficient-level trends remain physically consistent under the compact formulation. The wall-coefficient vector $\boldsymbol{\omega}$ assigns approximately $7.02\,\mathrm{dB}$ to brick/concrete and $1.46\,\mathrm{dB}$ to wood, closely matching COST-231 multi-wall expectations \cite{obiriStatisticalEvaluationIndoor2025a}. Within the environmental coefficient vector $\boldsymbol{\varepsilon}$, the dominant entries are negative overall, with $\varepsilon_{RH}\approx -0.082\,\mathrm{dB/\%}$ and $\varepsilon_{T}\approx -0.102\,\mathrm{dB/^\circ C}$, indicating net attenuation relief under typical HVAC and occupancy cycles. The packet-level SNR term remains the strongest single auxiliary predictor through $k_{\gamma}$. After Kalman prefiltering, however, the model relies less on instantaneous link-quality fluctuations: the SNR coefficient contracts in magnitude (from a fitted value of $-2.085\,\mathrm{dB}/\mathrm{dB}$ to $-0.372\,\mathrm{dB}/\mathrm{dB}$ after filtering), while the fitted path loss exponent stays within the expected indoor range, indicating improved separation between persistent propagation structure and transient disturbances.

\begin{figure}[hbt!]
    \centering
    \centerline{\includegraphics[width=0.8\linewidth]{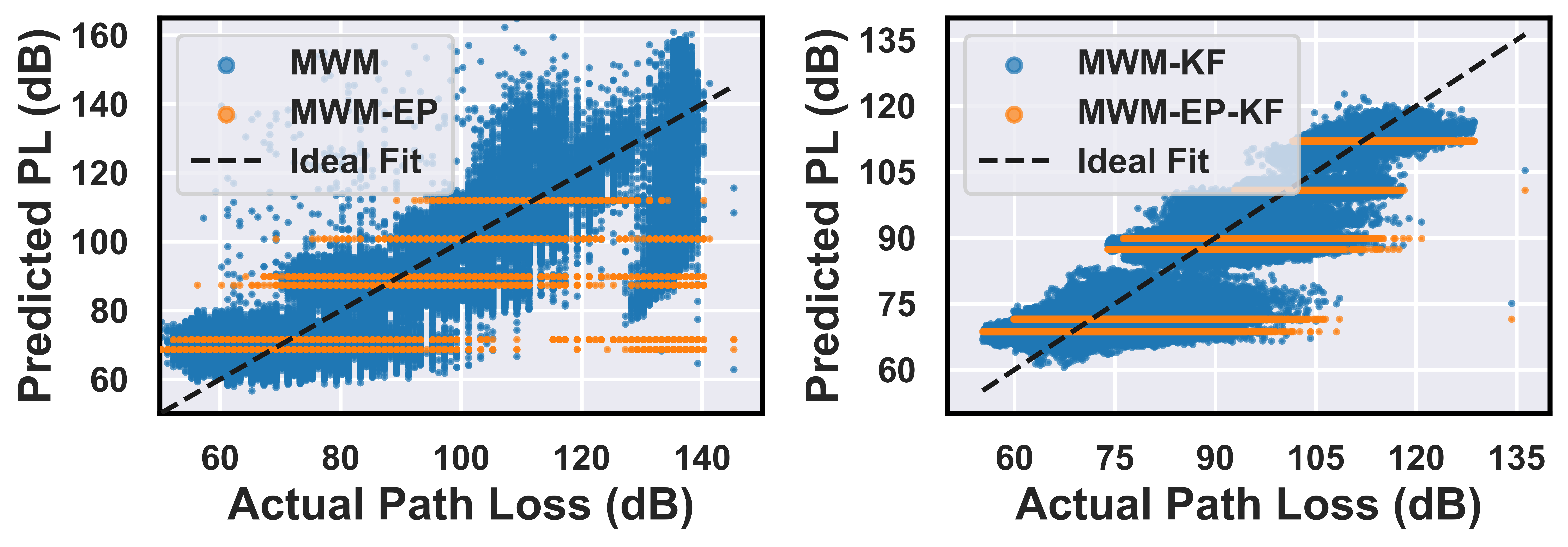}}
    \caption{\label{fig:residuals_kalman} Comparison of predicted and actual path loss for the MWM, MWM-EP, and MWM-EP-KF models with measured (left) and filtered RSSI (right).}
\end{figure}

Moreover, Kalman filtering compresses the heteroskedastic spread toward the identity line, resulting in tighter clusters and fewer pedestal bands (see Fig.~\ref{fig:residuals_kalman}). Quantitatively, the environment-augmented baseline reduces test RMSE from $\mathrm{10.95\,dB}$ to $\mathrm{8.09\,dB}$ while adding the Kalman prefilter reduces it further to $\mathrm{5.35\,dB}$ and raises $R^2$ to $0.89$, up from $0.82$ for the MWM-EP. Taken together, the environmental augmentation and Kalman prefiltering almost halve the residual RMSE and lift $R^2$, indicating that both systematic bias and shadowing variance are significantly reduced.
\begin{figure}[hbt!]
    \centering
    \includegraphics[width=\linewidth]{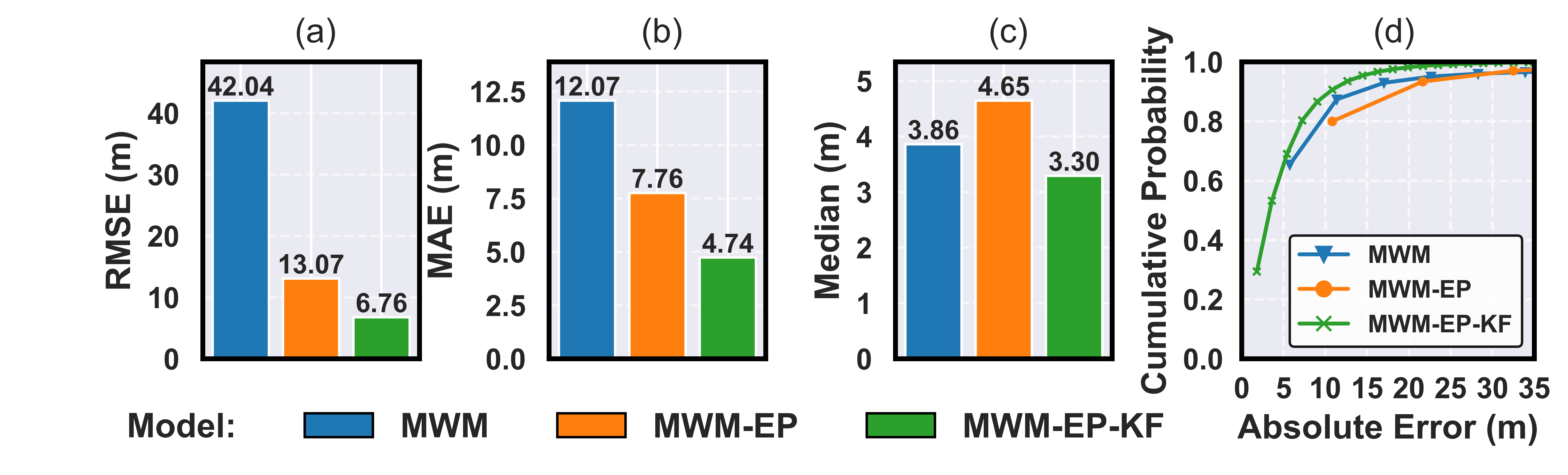}
    \caption{\label{fig:dist_metrics_cde} Distance estimation error comparison: (a) RMSE, (b) MAE, (c) median absolute error, and (d) empirical cumulative distribution error.}

\end{figure}

\subsection{Distance Estimation Analysis}

Precise indoor single-anchor ranging using LoRaWAN demands robust isolation of deterministic path loss from transient RSSI fluctuations. A hybrid approach coupling Kalman-filtered RSSI measurements with the MWM-EP model achieves superior distance estimation with an MAE of $\mathrm{4.74\,m}$ and an RMSE of $\mathrm{6.76\,m}$ (see Fig.~\ref{fig:dist_metrics_cde}(b)). This yields the lowest aggregate ranging error among the three evaluated variants. Environmental terms cut systematic attenuation by $26.13\%$, which is consistent with the RMSE gains reported for environment-aware combined path loss and shadowing modeling in \cite{obiriEnvironmentAwareIndoorLoRaWAN2025}, while the Kalman prefilter stabilizes RSSI \cite{bultenKalmanFiltersExplained2015}. Consistently, the cumulative distribution of absolute error (CDE) in Fig.~\ref{fig:dist_metrics_cde}(d) shows that $92.5\%$ of MWM-EP-KF estimates fall within $12\,\mathrm{m}$, compared to $88.0\%$ for MWM and $81.4\%$ for MWM-EP, confirming that the filtered, environment-aware model yields the highest fraction of low-error ranges. This ordering is also preserved at the device level: MWM-EP-KF yields lower mean and median absolute distance error than both MWM and MWM-EP on all six links (exact Wilcoxon signed-rank test on device-level summaries, $p=0.03125$ in all paired comparisons).

\begin{figure}[hbt!]
    \centering
    \centerline{\includegraphics[width=\linewidth]{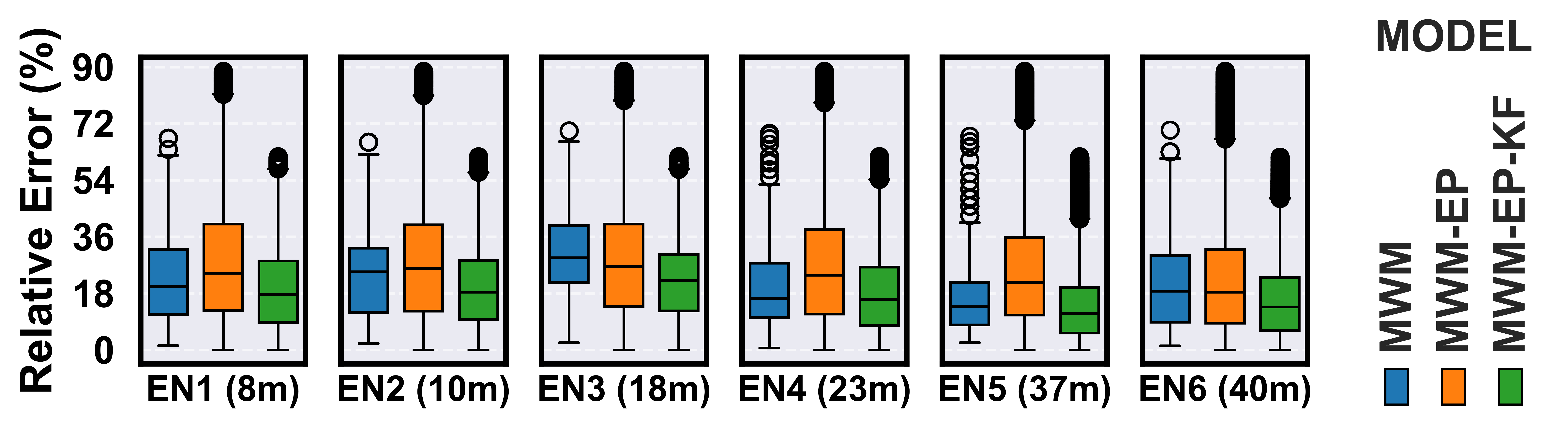}}
    \caption{\label{fig:relative_errors_boxplots} Relative error distributions per device for the MWM, the MWM-EP, and the MWM-EP-KF. Whiskers extend to $1.5\times$ the interquartile range.}
\end{figure}

Figure~\ref{fig:relative_errors_boxplots} indicates that MWM-EP-KF attains the lowest aggregate mean relative error across the six devices. Although some per-device interquartile ranges overlap, paired device-level summaries confirm that the filtered environment-aware model improves relative error over its unfiltered counterpart: MWM-EP-KF yields lower mean and median relative error than MWM-EP on all six links (exact Wilcoxon signed-rank test on device-level summaries, $p=0.03125$ in both cases). The gain is therefore not only visible in the compression of the error tails, but also supported by consistent link-level aggregates. At the aggregate level, the baseline MWM yields a mean relative error of $21.13\%$, whereas adding environmental parameters without filtering (MWM-EP) increases it to $26.18\%$, consistent with residual sensitivity to unfiltered sensor noise. Integrating Kalman filtering cuts this down to $18.11\%$. Gains are substantial per device: for instance, EN3's error drops from $32.45\%$ to $27.65\%$, and EN5's from $25.06\%$ to $16.95\%$. Unlike most ML solutions, the proposed MWM-EP-KF remains interpretable, enabling operators to pinpoint specific attenuation sources such as humidity or wall structures for targeted optimization in real-world IoT deployments.

\section{Conclusion and Future Outlook}
We presented MWM-EP-KF, an interpretable single-gateway ranging pipeline that couples an environment-aware multi-wall path loss model with a forward-only, innovation-driven Kalman prefilter for RSSI. On a one-year single-gateway office dataset, it achieves $4.74\,\mathrm{m}$ MAE, outperforming both a structure-only COST-231 multi-wall baseline ($12.07\,\mathrm{m}$) and its environment-augmented variant without filtering ($7.76\,\mathrm{m}$). Filtering reduces RSSI volatility ($\sigma$) from $10.33$ to $5.43\,\mathrm{dB}$ and the distance estimation RMSE from $42.04$ to $6.76\,\mathrm{m}$, while environmental augmentation cuts systematic error by $26.13\%$. These gains, achieved with $O(1)$ per-packet computation and physics-grounded coefficients, position MWM-EP-KF as a lightweight, interpretable ranging benchmark rather than a replacement for data-hungry fingerprinting or transfer-based localization pipelines. Taken together, they provide a practical single-anchor building block for future multi-gateway indoor LoRaWAN localization.

Some limitations remain. Our evidence comes from a single academic building with static nodes and one gateway, so performance may differ in settings with other architectures, materials, HVAC regimes, occupancy patterns, or mobility (including orientation changes and hand or body blockage). Our approach also assumes timely and accurate environmental sensing; sensor delays or drift can degrade estimates, and the filter's heuristically chosen noise levels may need to be retuned under different interference profiles. Although we have characterized rapid stabilization of the filtering stage, transfer to structurally new links or materially different buildings remains to be evaluated explicitly. Looking ahead, we will \textbf{(i)} validate across diverse sites (industrial, healthcare, warehouse) and with mobile nodes; \textbf{(ii)} extend to multi-gateway deployments (time alignment, diversity combining, distributed filtering); \textbf{(iii)} investigate online parameter adaptation via Extended or Unscented Kalman filtering, Bayesian evidence maximization, or reinforcement learning (RL) based auto-tuning; and \textbf{(iv)} explore multi-modal fusion (e.g., Bluetooth Low Energy (BLE) beacons, occupancy counters) to better resolve shadowing effects.

\bibliographystyle{IEEEtran}
\bibliography{refs}

\end{document}